# Frustration, dynamics and catalysis

R. Gonzalo Parra [1] and Diego U. Ferreiro [2,3*]

1. Life Sciences Department, Barcelona Supercomputing Center, Barcelona, Spain.
2. Protein Physiology Lab, Departamento de Química Biológica, Facultad de Ciencias Exactas y Naturales, Universidad de Buenos Aires, Buenos Aires C1428EGA, Argentina.
3. Instituto de Química Biológica de la Facultad de Ciencias Exactas y Naturales, Consejo Nacional de Investigaciones Científicas y Técnicas - Universidad de Buenos Aires, Buenos Aires C1428EGA, Argentina

* Corresponding author: ferreiro@qb.fcen.uba.ar

**Highlights**

- Local frustration sculpts proteins' functional energy landscapes

- Precise protein dynamics is essential for enzymatic activity

- Perturbations to local frustration leads to altered dynamics and activities

**Abstract**

The controlled dissipation of chemical potentials is the fundamental way cells make a living. Enzyme-mediated catalysis allows the various transformations to proceed at biologically relevant rates with remarkable precision and efficiency. Theory, experiments and computational studies coincide to show that local frustration is a useful concept to relate protein dynamics with catalytic power. Local frustration gives rise to the asperities of the energy landscapes that can harness the thermal fluctuations to guide the functional protein motions. We review here recent advances into these relationships from various fields of protein science. The biologically relevant dynamics is tuned by the evolution of protein sequences that modulate the local frustration patterns to near optimal values.

**Introduction**

The first proposal of the origins of the catalytic power of enzymes came almost 100 years ago, when Haldane posited that enzymes stabilize the transition state of the catalyzed reaction thus accelerating the chemical transformation step [1]. Detailed structural studies indeed showed that enzymes have specific regions where catalysis occurs and appropriately called them 'active sites'. However, the static picture of macromolecular objects fails to explain the origins of the catalytic power, and it has long been recognized that motions in proteins are essential for their biological roles [2–4]. Of course, molecules at 300K cannot be static objects but, as Hans Frauenfelder asserted, the point is to distinguish the Functionally Important Motions (FIMs) from the Biologically Unimportant Motions (BUMs) [5,6]. The FIMs may facilitate substrate binding and alignment in the reaction center, barrier crossing, product release, etc, and they must underlie the secrets of allostery, the necessary condition that protein activity must be physiologically adjustable by external regulators.

**Local frustration tunes enzymes' internal dynamics**

Today we understand protein motions in the language of the Energy Landscape Theory [7]. This theory is based on the appreciation that natural proteins are evolved systems for which there is a fundamental connection between the energy distributions and the structural populations [8]. The largest protein movements are related to the overall folding of the polypeptide chain to their native states, which is possible for systems that follow the Principle of Minimal Frustration [9], and have associated landscapes that resemble a rough funnel [10] (Figure 1). The funnel is mainly given by the parts that reinforce each other fitting well together and the roughness is brought about by the conflicting interactions. Minimal frustration however does not rule out that some local frustration may be present in folded biomolecules. Moreover, it is becoming clearer that the remaining frustration is not just a random accident but plays an essential part of the inner workings of protein molecules [11,12]. Local frustration gives rise to the asperities of the landscape at the bottom of the funnel, which is related to the population of the conformational substates (CS) and their interconversion [13] (Figure 1).

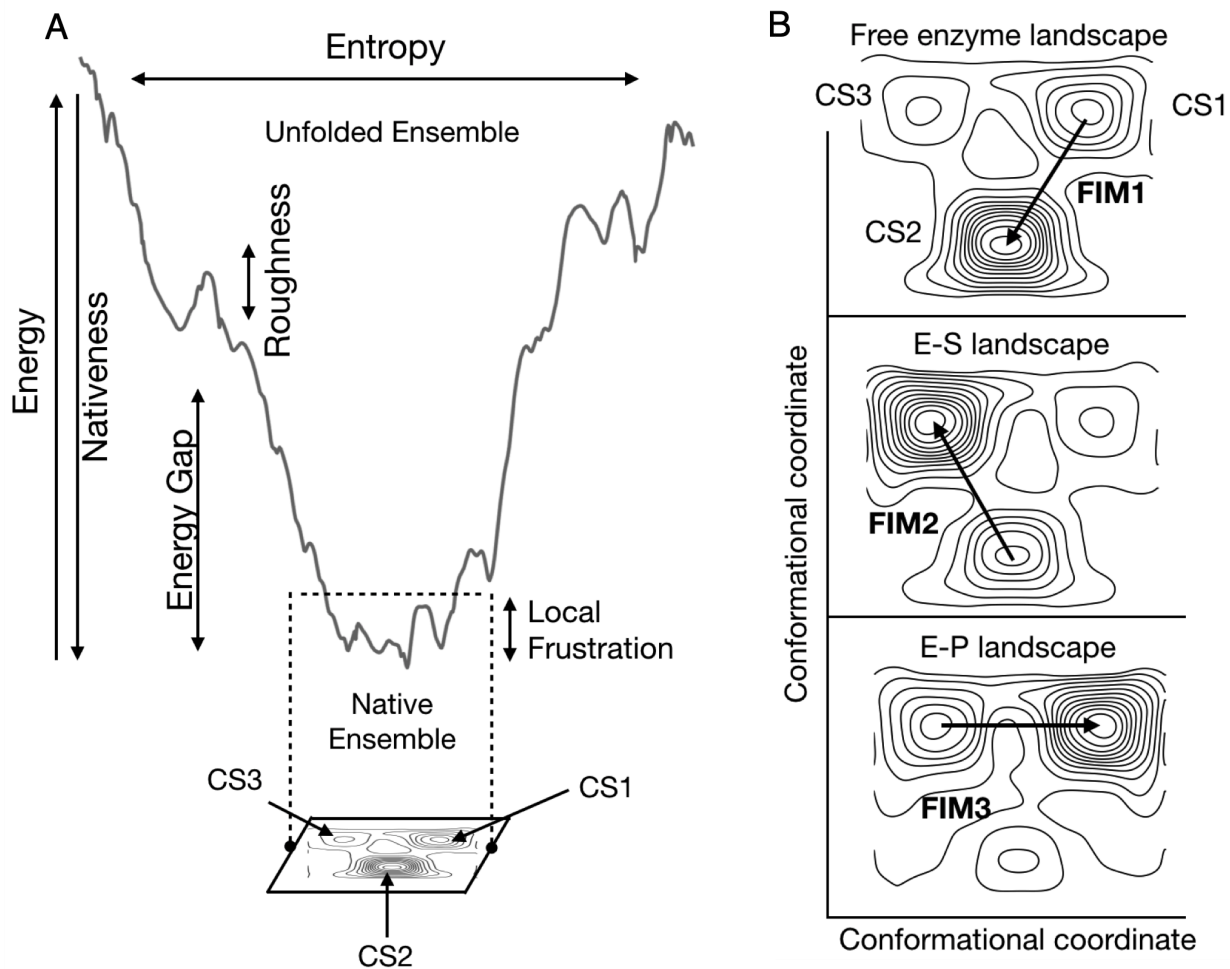

**Figure 1:** Enzymes explore dynamic energy landscapes. A) the overall topography of a globular protein energy landscape is pictured [10]. The unfolded ensemble with high energy and high entropy is at the top of the funnel and the native ensemble with low entropy and low energy at the bottom. Local frustration shapes the landscape's ruggedness, defining the Conformational Substates (CS) and their population. In this illustrative case, three CS can be populated, with the free energy contour levels shown. B). Contour plots of the conformational free energy landscapes corresponding to three stages of a simple catalytic cycle: the free enzyme (top), enzyme–substrate complex (middle), and enzyme–product complex (bottom). The free energy of the CSs changes upon binding substrate (E-S) or product (E-P) and the Functional Important Motions (FIMs) drive the transitions between CSs [14].

We should bear in mind that the energetic differences here are given by a myriad of weak non-covalent interactions between the amino-acids' atoms, the solvent and the reacting chemical species, so one could anticipate great physical complexity. However, the fact that many experimental studies of enzyme-catalyzed reactions can be well-approximated with pseudochemical reaction coordinates and transition state theory reveals that there are emergent collective motions that can lump together thermodynamics states and define effective barriers between them (Figure 2). Too little frustration freezes the FIMs because the interactions are too strong and some CS dominates the population. Too much frustration and many CS become available, so the BUMs dominate the dynamics. A sweet spot of "just right" local frustration allows for the emergence of the FIMs that optimize function (Figure 2). The biologically relevant dynamics is tuned by evolution of protein sequences that may even give rise to 'perfect enzymes' as Albery and Knowles foresaw [15] or at least to good enough ones [16].

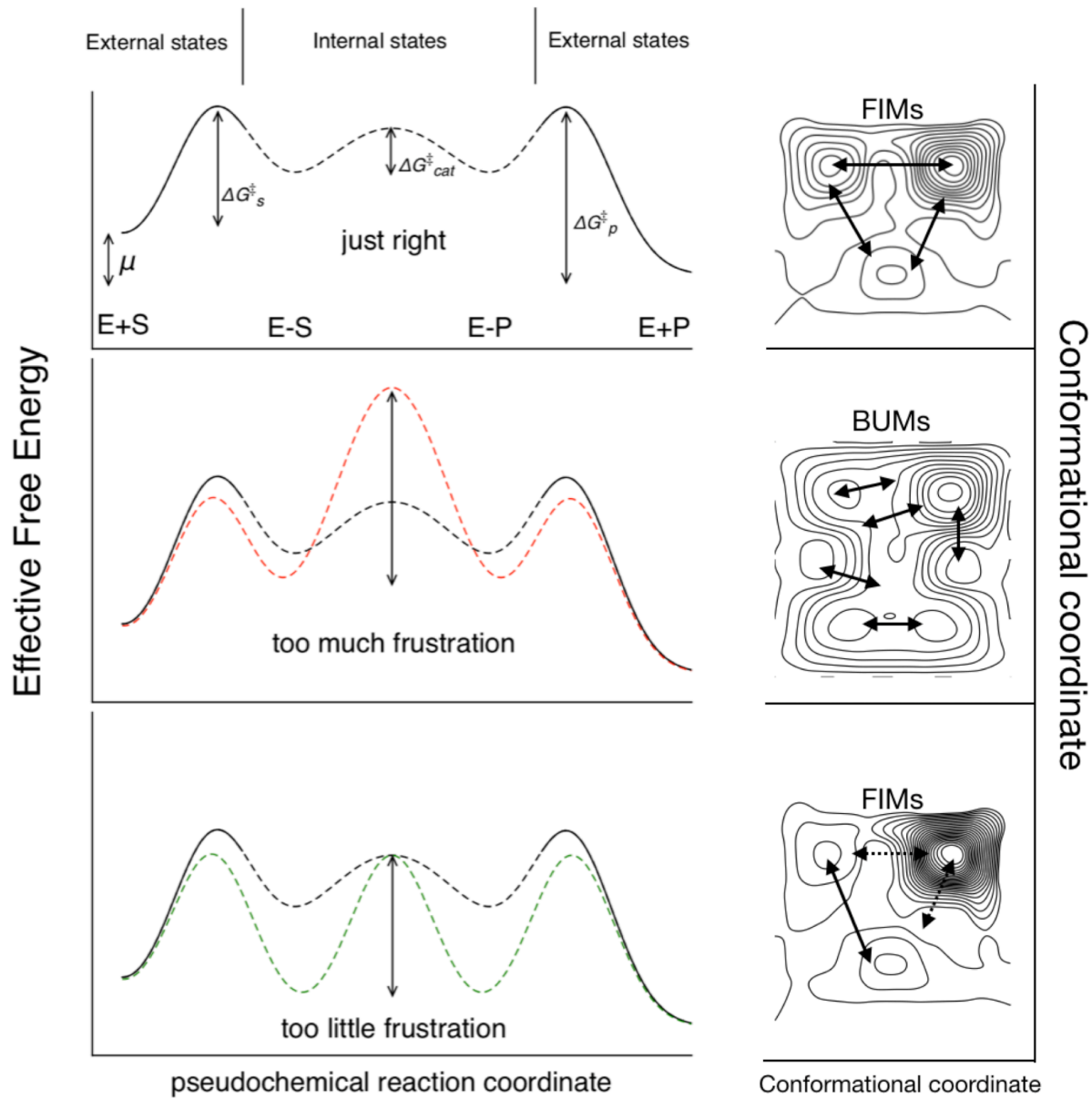

**Figure 2:** Cartoon of the pseudochemical reaction pathway of a simple enzyme. The system is divided into external (i.e. free enzyme) and internal collective states (i.e. the enzyme-substrate and enzyme-product complexes). The structure and dynamics of the internal states give rise to effective free energy barriers ΔG$^{‡}$. Local frustration modulates the internal energetics and as such the population of the structural forms and their interconversion (right panels). Too little frustration (bottom) freezes the relevant protein motions and the effective barriers increase, slowing down catalysis. Too much frustration (middle) and the relevant motions are less probable, slowing down catalysis. A sweet spot of "just right" local frustration allows for the emergence of the relevant dynamics that optimize function (top).

Today, there is no definitive way to determine and quantify the local frustration in atomically complicated objects like natural proteins. However, a simple heuristics based on energy landscape theory has been shown to be very useful for analyzing how the stabilization energy in protein domains is distributed [17,18]. In 2019 Freiberger et al surveyed all experimentally annotated natural enzymes in the Catalytic Site Atlas (CSA) and found, in agreement with previous hypotheses, that the catalytic sites themselves are enriched in highly frustrated interactions, regardless of the protein oligomeric state, overall topology, and enzymatic class [19]. A secondary shell of more weakly frustrated interactions typically surrounds the catalytic site itself. When various family members of major enzyme classes were analyzed, it was found that the energetic signatures are more evolutionarily conserved than the primary structure. It is apparent that, at least at a coarse grain level, local frustration presence is a general aspect of natural enzymes. A gallery of local frustration patterns is shown in Figure 3. In these 'Frustratograms' the red lines indicate pairwise contacts between residues where the local energetics are not folding-optimized as compared to the distribution of decoys and mark highly frustrated interactions. These interactions typically patch at the surface of globules, account for ~10% of all interactions and are often associated with functional regions [11, 17]. Green lines indicate contacts with folding-optimized energetics as compared to decoys, marking minimally frustrated contacts, typically comprising ~40% of the interactions and commonly related to stability and foldability. The third type of interaction, the neutral contacts, whose native energy is statistically indistinguishable from the mean energy of random decoys is not drawn and typically represents ~50% of all contacts.

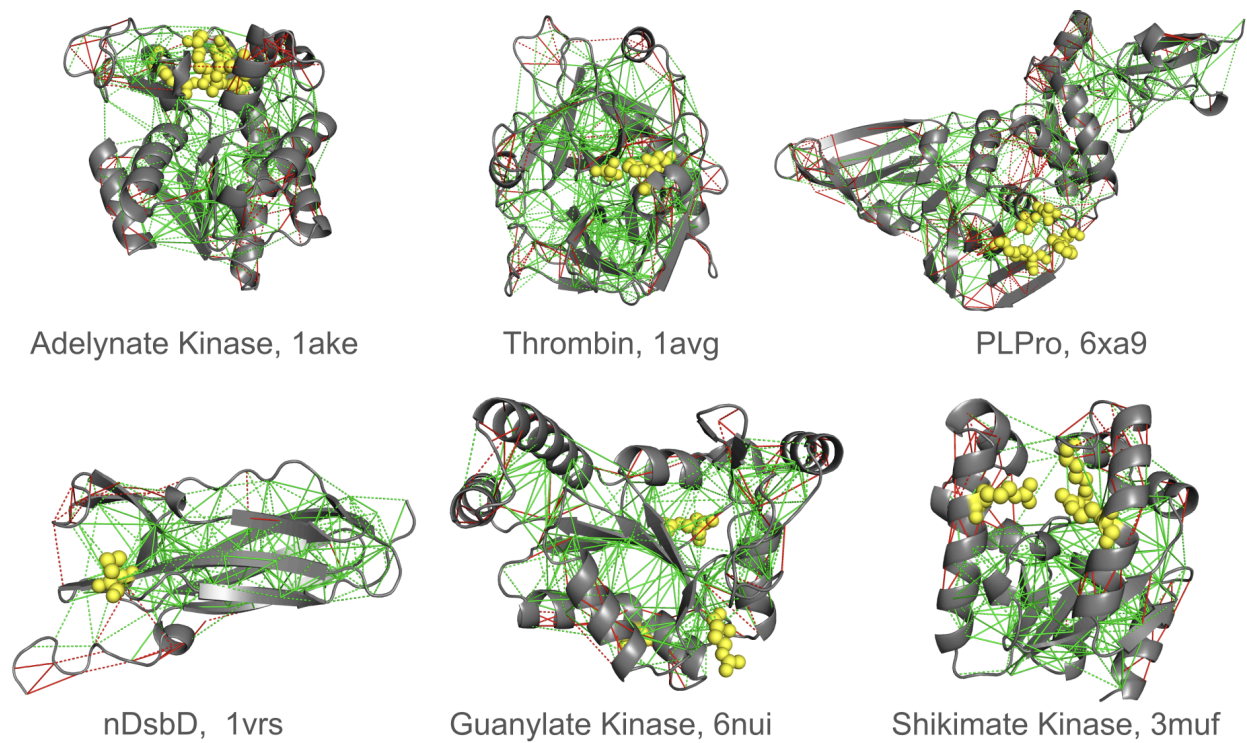

**Figure 3.** Frustratograms of some of the enzymes discussed in this article. The structures represent the protein backbone in gray ribbons and the catalytic sites in yellow balls. Pairwise residue contacts are colored according to their configurational local energetic frustration index calculated by the coarse-grained frustratometeR tool [18]. Red lines indicate highly frustrated contacts, and green lines indicate minimally frustrated interactions. Regions of high local frustration co-localize with active-sites, yet other patches of highly frustrated interactions are present, presumably for allostery [19]. The enzymes shown are (top row, left to right), common name and (pdb code): Adenylate Kinase (1ake), Thrombin (1avg), SARS-CoV-2 papain-like protease (PLPro, 6xa9); (bottom row, left to right): nDsbD (1vrs), Guanylate Kinase (6nui), and Shikimate Kinase (3muf).

Local frustration can act as an evolutionary fingerprint of functional adaptation. In SARS-CoV-2 PLPro, a highly frustrated residue (Trp106) and two uniquely highly frustrated residues (Thr225 and Lys232) at the ISG15-binding site emerged recently in the Betacoronavirus phylogeny, suggesting a functional innovation that may underlie altered substrate specificity and host interaction preferences [20].

A recent large-scale study by Hou et al. [21] supports the idea of catalytic residues being inherently associated with structural destabilization, as expected for highly frustrated interactions. Using a statistical potential-based approach (SWOTein), the authors analyzed hundreds of enzymes and showed that catalytic sites are consistently located in regions of stability weakness, which are surrounded by shells of residues that are comparatively more stabilizing. Intriguingly, they found an oscillatory pattern of alternating stability strengths and weaknesses that diminishes with distance from the active site that is observed across enzyme classes, regardless of solvent accessibility. In agreement with this, the destabilizing catalytic residues are also among the most evolutionarily conserved, while surrounding compensatory shells display lower conservation.

A common experimental realization that local frustration is at play in enzymes is the often found case that mutations at catalytic sites often make protein folds more stable, while destabilizing mutations elsewhere are the norm [22]. Of course, mutations at catalytic sites often make the enzyme less catalytically efficient and these are usually referred to as stability-activity tradeoffs [23]. It is important to recall that mutations far away from the active site are often found to impact catalytic efficiency [24]. The effect of such mutations can only be understood in terms of the dynamics of the CS populations along the catalytic cycle.

**Functionally important motions drive catalysis**

Functionally important motions (FIMs) in enzymes take place across a range of timescales, which can be modulated by local frustration. The study of the motions of the protease Thrombin, in the microsecond to millisecond timescale, revealed a striking correlation between the presence of frustrated interactions and regions undergoing slow time scale dynamics [25]. Recent work by Weinreb et al. [26] provides enthralling experimental support for the idea that functionally important enzyme motions operate under mechanical constraints. By probing the enzyme Guanylate Kinase using nano-rheology, which enables studying protein motions on timescales of 10 to 200 milliseconds, the authors reveal that single-point mutations in high-strain regions, identified from conformational transitions and AlphaFold predictions, can significantly dampen both protein flexibility and catalytic activity [26]. These frustrated high-strain regions correspond to sites that deform during substrate binding, and can be interpreted as viscoelastic domains: regions that behave both like elastic solids (storing mechanical energy) and like viscous fluids (dissipating energy as internal friction). This dual mechanical nature reflects a need for dynamic adaptability during the catalytic cycle. Crucially, these mechanically soft, deformable zones often lie far from the active site, yet still influence catalysis through long-range dynamical coupling.

Adenylate kinase (AdK) is a remarkable system for which the coupling of motion with catalysis has been investigated. This protein undergoes a large structural transition during the catalytic cycle, and 'cracks' during the transition [27]. Large-scale motions of biomolecules involve linear elastic deformations along low-frequency normal modes, but for function nonlinearity is essential. Local unfolding, or cracking, occurs in regions where the elastic stress becomes too high during the transition, and if this region is enriched in frustrated interactions, the transition is favoured [28]. The concept of cracking explains the surprising experimental observation that many enzymatic activities are enhanced with low concentration of denaturants. A recent study on P-ATPases shows that this effect is related to the local frustration changes along the catalytic cycle [29].

Another study on AdK has shown that steric frustration, arising from spatial incompatibility between a newly bound substrate and an incompletely released product, can actively facilitate product dissociation, overcoming a rate-limiting bottleneck in the enzymatic cycle [30]. Molecular dynamics simulations revealed that when ATP binds to AdK before ADP has fully dissociated, steric clashes between the phosphate groups destabilize the ADP binding pocket, reducing the energy required for its release. This introduces a novel substrate-driven mechanism for product expulsion, challenging the traditional view that product release is a purely passive process governed by affinity differences. A follow-up study elucidated the underlying physical interactions that drive this frustration-mediated acceleration [31]. Beyond steric clashes, electrostatic repulsion between the negatively charged phosphate groups of ATP and ADP also plays a significant role in destabilizing product binding. These interactions disrupt a key hydrogen bond network at the active site which normally stabilizes the closed conformation of the enzyme, lowering the free energy barrier for domain opening from ~20 kcal/mol to ~13 kcal/mol in the presence of ATP.

The concept of frustration-driven catalysis extends beyond AdK and may represent a general principle in multi-substrate enzymes. Unlike ground-state destabilization models, where binding affinity is reduced to facilitate release [32], frustration actively creates a dynamic imbalance that promotes conformational transitions and accelerates product dissociation. Similar mechanisms may exist in other enzymatic systems where product release is rate-limiting, such as ATPases and kinases, where substrate binding could generate local frustration to drive turnover [4].

Recent work by Burns et al. [33] provides an example of how local frustration in flexible, disordered loops can be dynamically modulated to fine-tune enzymatic activity. Focusing on homologs of the C-terminal domain of bacterial enzyme I (EIC), they identified temperature-sensitive residue contacts within disordered catalytic loops that change predictably with temperature and directly influence catalytic efficiency. Using Hamiltonian replica exchange molecular dynamics and mutagenesis, they demonstrated that mutations to the most temperature-sensitive residues enhance activity at suboptimal temperatures, while mutations to insensitive sites have little effect. This suggests that local frustration around active sites is not only statically encoded but can also be actively modulated via sequence-dependent alterations in loop dynamics.

A striking case of redox-modulated local frustration is illustrated by Stelzl et al. [34], who examined the bacterial oxidoreductase nDsbD. They showed that the formation of a disulfide bond in the oxidized form of the enzyme introduces local frustration near the active site, disrupting the packing of a cap-loop that otherwise shields catalytic cysteines. This frustration enables the loop to adopt open conformations in the absence of a binding partner, thereby facilitating a conformational selection mechanism critical for downstream electron transfer. In contrast, the reduced form lacking the disulfide remains minimally frustrated, with the closed loop protecting the thiols from oxidation. This redox-dependent switch exemplifies how local frustration can be dynamically regulated to control loop flexibility and enzymatic function.

Long-range electron transfer in metalloproteins relies on an evolutionary refined interplay of donor-acceptor electronic coupling, electrochemical potentials, and active-site reorganization energies. Chen et al. [35] highlighted this in the Azurin enzyme, where a precisely orchestrated set of local and distant interactions in the wild-type protein minimizes energetic frustration, enabling fast electron flow. Conversely, engineered azurin variants with sluggish electron transfer display elevated energetic frustration both locally and distantly from the copper site, and this frustration is more strongly influenced by the copper's oxidation state.

Another illustration of how local frustration modulates enzymatic function comes from recent molecular dynamics studies of Shikimate Kinase (SK) by Ojeda-May [36]. In this work, mutations at residue R116, located near the active site, were shown to significantly alter the global conformational dynamics of the holoenzyme. The R116A and R116K mutants exhibited increased local frustration in the vicinity of the mutation site, which propagated to long-range structural perturbations affecting domain coupling and motion. These changes correlated with experimentally observed reductions in catalytic efficiency, highlighting the residue's role in coordinating both catalysis and product release. The findings support the hypothesis that highly frustrated residues act as functional control points, where local energetic imbalances enable the conformational changes required for multi-step enzymatic turnover.

**Design Implications and Future Perspectives**

Massive machine learning is beginning to be a useful tool to explore the connections between local frustration, dynamics and function. Recent work to redesign the Tobacco Etch Virus (TEV) protease highlights the necessity of explicitly preserving functional residues during structure-based sequence optimization. Sumida et al. [37] generated unconstrained designs by reverse-folding sequences with ProteinMPNN [38] to the TEV backbone without fixing any residues. These variants exhibited significantly improved expression and thermal stability but were catalytically inactive, as ProteinMPNN replaced functional residues with alternatives optimized for structural integrity. To preserve activity, the authors show that they needed to fix not only catalytic residues but at least 50% of the most conserved residues on a multiple sequence alignment of TEV homologs. Even more, some of these designs showed up to 26-fold increases in catalytic efficiency compared to the native enzyme. These results emphasize that functionally essential sites often coincide with local energetic frustration and are therefore systematically disfavored by stability-driven design models. The need to fix a substantial proportion of residues, not in the active site, to preserve activity suggests that ProteinMPNN was replacing positions important for dynamics and allosteric communication with more stabilizing alternatives, thereby impairing function. Exploring the conservation of frustration at the family level of proteins [20] could be used as a tool to identify highly frustrated and functionally relevant residues to advise which residues to be fixed in order to avoid loss of function in ProteinMPNN designs.

The concept that proteins must move to perform biological functions is not new as the motions of certain protein residues could contribute to catalysis by facilitating access to the transition state [39]. The energy landscape perspective provides us with concepts and tools to explore the connections between local frustration, CS population, FIMs and chemical reaction. Recently, Guan et al. have shown how frustration analysis can be used to empower machine learning algorithms such as AlphaFold2 to generate alternate structures along allosteric pathways, focusing on the AdK landscape [40]. However, we are still lacking a quantitative general mechanism that allows us to design proteins that catalyse reactions with the precision and efficiency of natural enzymes. If frustration-driven mechanisms can be identified in other enzyme families, they could be harnessed to design biocatalysts with enhanced turnover rates or to develop inhibitors that exploit dynamic or steric incompatibilities to selectively slow down enzymatic activity. Evolution may have explored such lower and upper bound frustration limits as shown for the two allelic forms of the same enzyme coexist in human populations, each of them showing different degrees of local frustration which may confer adaptive benefits to different living environments [41]. Future research could determine how frustration-mediated acceleration strategies vary across different classes of enzymes, potentially uncovering universal principles of enzymatic activity. The growing recognition of local frustration as a functional rather than detrimental feature in proteins provides a new framework for understanding enzyme evolution, dynamics, and design.

## Acknowledgements

The authors thank Peter G. Wolynes for the insightful discussions and suggestions on the manuscript, and Ezequiel Galpern for his help with Figure 1. RGP was supported by a fellowship from Grant IHMC22/00007/ PRTR funded by the Instituto de Salud Carlos III (ISCIII) and is currently funded by a Ramon y Cajal grant (RYC2023-043825-I). DUF was supported by the Consejo de Investigaciones Científicas y Técnicas (CONICET); CONICET Grant PIP2022-2024—11220210100704CO and Universidad de Buenos Aires grant UBACyT 20020220200106BA. We call the attention of the international scientific community about the catastrophic erosion of Argentina's strong scientific tradition due to current funding constraints and the sudden termination of long term policies.